\documentclass[aip,jcp,amsmath,amssymb,reprint]{revtex4-1}
\usepackage{dcolumn}
\usepackage{bm}
\usepackage{graphicx}

\begin{document}
\preprint{AIP/123-QED}
\title[Single-File Diffusion in an Interval:  First Passage Properties]
{Single-File Diffusion in an Interval: First Passage Properties}
\author{Artem Ryabov}
\email{rjabov.a@gmail.com}
\affiliation{Charles University in Prague, Faculty of Mathematics and Physics, Department of Macromolecular Physics, V Hole{\v s}ovi{\v c}k{\' a}ch 2, 180~00~Praha, Czech Republic} 
\date{\today}
\begin{abstract}
We investigate the long-time behavior of the survival probability of a tagged particle in a single-file diffusion in a finite interval. The boundary conditions are of two types: 1) one boundary is absorbing the second is reflecting, 2) both boundaries are absorbing. For each type of the boundary conditions we consider two types of initial conditions: a) initial number of particles $N$ is given, b) initial concentration of particles is given ($N$ is random). In all four cases the tagged-particle survival probability exhibits different asymptotic behavior. When the both boundaries are absorbing we also consider a case of a random interval length (single-file diffusion on a line with randomly distributed traps). In the latter setting, the initial concentration of particles has the same effect on the asymptotic decay of the survival probability as the concentration of traps. 
\end{abstract} 
\maketitle 
\section{Introduction}  

The diffusion of particles in narrow channels where the particles cannot pass each other is known as the single-file diffusion (SFD). Although 
the overall density of particles in such a system evolves as if there is no interaction,\cite{Redner2} the diffusion of one specific particle (a tagged particle, or a tracer) is slowed down by the exclusion interactions with neighboring particles. Experimentally SFD is observed for instance  in narrow biological pores,\cite{Hodgkin} inside zeolites,\cite{HahnKarger, KargerBOOK} during the sliding of proteins along DNA,\cite{DNAnature, Hippel} and in many other either natural or artificial systems\cite{Halpin, nanoWaterI, nanoWaterII, Superionic, SuperionicII, SSnanopores, BechCircle, BechPhoto, colloidpolym, ChargedSpheres, BallsOFsteel}. 

Theoretical description of the dynamics of tagged-particles in SFD was pioneered by Harris.\cite{Harris} He showed that the mean-square displacement of the tagged particle grows with time as $t^{1/2}$ (in contrast to the linear increase for the normal diffusion). After that many authors studied  the tagged-particle dynamics under different conditions (for a comprehensive review see Introduction in Ref.\ \onlinecite{BOX2}). 
Despite the long history of the problem, studies devoted to SFD in finite systems appears only recently.\cite{BOX1, BOX2, BOX3, BOX4} Most of these  studies have focused on closed systems where the number of particles is conserved. An exception is Ref.\ \onlinecite{Rodenbeck} where the tagged-particle propagator for a finite interval with two absorbing boundaries was discussed as a particular application of the general formalism. In the present paper we go beyond the previous studies in the followings: 1) we consider SFD in a finite interval with absorbing boundaries, 2) instead of the moments of tagged-particle probability density function we study the long-time behavior of the tagged-particle survival probability. 

The absorbing boundary conditions can be used to model \emph{traps}, and  the trapping model (capturing of diffusing particles by immobile traps) represents the simplest model of diffusion-limited reactions\cite{Havlin, HavlinAdvances, Redner} which occur in many real systems (examples of such systems are cited in the first paragraph). Up to our best knowledge, there are only a few studies of the effect of hard-core interaction on the first passage properties. For a recent study of a first passage time density of a tagged particle in an infinite system see Ref. \onlinecite{FPTSFD}. There are also earlier works focusing on the overall absorption rate in systems with dissimilar interacting particles.\cite{Bunde1, Bunde2} The semi-infinite interval with the absorbing boundary at the origin was considered in Ref.\ \onlinecite{RC2012}, where the asymptotics of the tagged-particle survival probability was discussed. 

In the present paper we consider a diffusion of identical hard-core interacting Brownian particles in the one-dimensional interval of the length $L$. We discuss two types of initial conditions and two types of boundary conditions. 
As for the boundary conditions, we consider: 1) a perfectly absorbing boundary at $x=0$ and a reflecting boundary at $x=L$, and 2) both boundaries are perfectly absorbing. 
As for the initial conditions, we discuss two settings: a) initially there are exactly $N$ particles homogeneously distributed within the interval; b) the initial concentration of particles $c$ is given and it is homogeneous within the interval. Hence, in the latter setting, the initial number of particles is a Poisson random variable with a mean value $cL$. In all four cases it is assumed that as soon as the particle hits an absorbing boundary, it is immediately removed from the system.  

The paper is organized as follows. In Sec.\ \ref{taggedpdyn}, we consider general boundary conditions. First, for a given initial number of particles $N$ (Sec.\ \ref{FixedNgeneral}), we present the probability density function (pdf) of a tagged particle's position (Eq.\ (\ref{taggedPDF})) and a survival probability of the tagged particle (Eq.\ (\ref{SnN})). The pdf (\ref{taggedPDF}) represents the starting point of all further analyses. Second, we derive the same two quantities for a given initial concentration $c$ (Sec.\ \ref{Fixedcgeneral}, Eqs.\ (\ref{pn}), (\ref{Sn})). Sec.\ \ref{Transmission} presents the asymptotics of the survival probability in the case when the left boundary is absorbing and the right boundary is reflecting (Eqs.\ (\ref{SnNT}), (\ref{SnNTe}) for a given $N$, Eq.\ (\ref{SnT}) for a given $c$). 
Sec.\ \ref{Absorption} contains the asymptotics of the survival probability in the case when both boundaries are absorbing (Eq.\ (\ref{SnNA}) for a given $N$, Eq.\ (\ref{SnA}) for a given $c$). Further, in Sec.\ \ref{randomintlength}, the survival probability is discussed when the interval length $L$ is random (cf. Eq.\ (\ref{SnAaveragedAsy})).  The paper is concluded by the interpretation of the results (Sec.\ \ref{concludingremarks}).

\section{Dynamics of a tagged particle: general boundary conditions} 
\label{taggedpdyn}
\subsection{Fixed initial number of particles} 
\label{FixedNgeneral}

The basic quantity which is assumed to be known is the pdf of a single noninteracting particle, $f(x,t)$. It satisfies the diffusion equation
\begin{equation}
\label{Diff}
\frac{\partial}{\partial t}f(x,t) = D \frac{\partial^2}{\partial x^2}f(x,t)\,\,,\quad x\in (0,L)\,\,,
\end{equation}
subject to appropriate boundary conditions at $x=0$ and $x=L$, and to the initial condition $f(x,0)=1/L$, $x\in (0,L)$. 

Assume that initially there are $N$ particles distributed randomly in $(0,L)$. Let us label the particles by numbers $1,\ldots,N$, from left to right. Due to the hard-core interaction the particle order is conserved for all times. For a given $x$, $x\in (0,L)$ the exact pdf of the tagged particle with number $n$, $n=1,\ldots,N$, reads 
\begin{eqnarray} 
\label{taggedPDF}
p_{n:N}(x,t) &=& \frac{N!}{(n-1)!(N-n)!} f(x,t)\\
\nonumber
& & \times \left(F(x,t)\right)^{n-1} \left(1-F(x,t)\right)^{N-n}, 
\end{eqnarray} 
where $F(x,t)$ is a probability of finding the single noninteracting particle to the left of $x$ (either in a left trap or diffusing within an interval $(0,x)$). We specify $F(x,t)$ in subsequent sections (cf. Eq. (\ref{CDF1}) when the right boundary is reflecting, and Eq. (\ref{CDF2}) when both boundaries are absorbing), the results of the present section are valid for a general $F(x,t)$. Pdf (\ref{taggedPDF}) constitutes the starting point for all subsequent analyses. 

\begin{figure}[t]
\begin{center}
\includegraphics[scale=0.35]{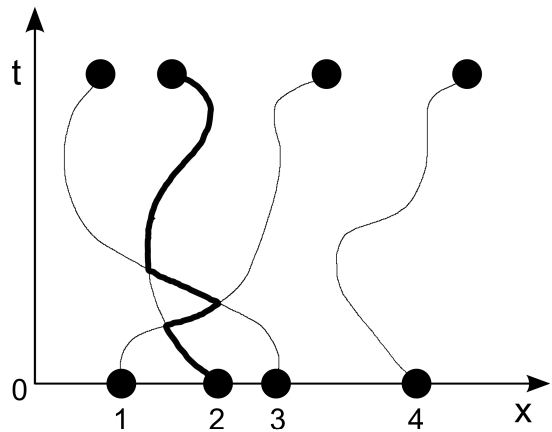}
\caption{Illustration of space-time trajectories of $N=4$ particles. During the time interval $(0,t)$, the tagged particle (bold line, $n=2$) undergoes three elastic collisions with its neighbors. Each collision can be equivalently represented as follows: we let two approaching particles pass freely through one another, and after they cross we exchange their labels. This maps the trajectories of interacting particles onto the trajectories of \emph{noninteracting} ones. Namely, the trajectory of the tagged particle is equivalently represented by the trajectory of the second left-most particle among $N=4$ noninteracting particles (regardless the actual identity of the second left-most noninteracting particle). The mapping implies that, at the time $t$, the pdf for the tagged particle equals to the pdf for the second left-most particle among $N=4$ noninteracting particles. For a general $n$, and $N$, the pdf is given in Eq.\ (\ref{taggedPDF}).
} 
\label{fig1}
\end{center}
\end{figure}

To justify Eq.\ (\ref{taggedPDF}) notice the following equivalence between the dynamics of hard-core interacting particles and of noninteracting ones. If two neighboring hard-core particles collide, they bounce off each other and their order is preserved. Equivalently, we can assume that, instead of the collision, the two particles pass freely (without interaction) through one another and after that we just exchange their labels. This is the main idea behind Eq.\ (\ref{taggedPDF}).
According to this idea the dynamics of interacting particle is generated merely by exchanging the labels of noninteracting particles whenever the two of them pass each other  (for an illustration, see Fig.\ \ref{fig1}). Therefore \emph{the pdf for the $n$-th tagged particle equals to the pdf for the $n$-th left-most particle among $N$ noninteracting ones}. The latter pdf is on the right-hand side of Eq.\ (\ref{taggedPDF}). To see this, assume that each of the noninteracting particles is described by pdf $f(x,t)$, then the right-hand side of Eq.\ (\ref{taggedPDF}) (multiplied by ${\rm d}x$)  yields the probability that there is a particle in $(x, x + {\rm d}x)$ and, simultaneously, there are precisely $n-1$ ($N-n$) particles to the left (to the right) of it. The factorials account for all possible sequences of particles to the left and to the right of the $n$-th particle. For a further discussion see Refs. \onlinecite{Redner2}, \onlinecite{Levitt}, or \onlinecite{Percus}. The pdf (\ref{taggedPDF}) can also be derived formally by solving a hierarchy of coupled diffusion equations as it was done in Ref. \onlinecite{RC2012} for a semi-infinite interval with the absorbing boundary at the origin (cf. Eq.\ (35) in Ref. \onlinecite{RC2012}).

The survival probability, that is the probability that, the tagged particle does not hit an absorbing boundary before $t$ is given by the spatial integral 
\begin{equation}
S_{n:N}(t) = \int_{0}^{L}\!\!\! {\rm d} x\, p_{n:N}(x,t)\,\,. 
\end{equation}
It reads (see Appendix \ref{AppendixS} for derivation)
\begin{eqnarray}
\label{SnN}
&& S_{n:N}(t) =  n  \binom{N}{n} \sum_{k=0}^{n-1}\frac{(-1)^{k}}{N-n+k+1} \binom{n-1}{k}
\times \\ 
\nonumber 
&&\,\, \times \left[ \left(1\!-\!F(0,t)\right)^{N-n+k+1}-  \left(1\!-\! F(L,t)\right)^{N-n+k+1} \right]\,\,. 
\end{eqnarray}

\subsection{Fixed initial concentration of particles}
\label{Fixedcgeneral}

The finite interval with a fixed initial concentration of particles $c$ can be modeled as follows. We imagine that, for $t<0$, the interval $(0,L)$ is a part of the whole real line on which the particles are distributed randomly with the uniform concentration $c$. Subsequently, at $t=0$, the interval is decoupled from the line by the boundaries at $x=0$ and $x=L$. During the diffusion $(t>0)$, if a particle hits an absorbing boundary, it leaves the interval and will never return. 

Initially there are on average $cL$ particles within the interval $(0,L)$. The probability that the initial number of particles equals $N$ is given by the Poisson distribution 
\begin{equation} 
\label{Poisson} 
P(N) = \frac{(c L)^{N}}{N!}\, {\rm e}^{- cL}\,\,.
\end{equation}
For a given $N$, the tagged particle pdf is given by Eq.\ (\ref{taggedPDF}).
In the present case, the pdf of the $n$-th tagged particle regardless the initial number of particles is obtained from pdf (\ref{taggedPDF}) by the summation   
\begin{equation}
p_{n}(x,t) = \sum_{N=n}^{\infty}  p_{n:N}(x,t) P(N)\,\,,
\end{equation} 
which yields  
\begin{eqnarray}
\label{pn} 
p_{n}(x,t) &=&  c L f(x,t) \frac{\left[c L F(x,t)\right]^{n-1}}{(n-1)!}\exp\left[-cL F(x,t)\right] \,\,. 
\end{eqnarray}

How to interpret pdf $p_{n}(x,t)$? Is it normalized to one at $t=0$? To give the answers let us return for a while to the fixed $N$ case. It is easy to show that 
\begin{equation}
\sum_{n=1}^{N}p_{n:N}(x,t) = N f(x,t)\,\,, \quad  \int_{0}^{L}\!\!\! {\rm d} x\, p_{n:N}(x,0)=1 \,\, .
\end{equation}  
The first equation manifests the well known fact that the overall density of particles evolves as if there is no interaction. The second equation is a normalization condition -- we assume that, at $t=0$, there are $N$ particles in the interval hence the $n$-th particle is obviously also there. Returning to the fixed $c$ (random $N$) case, the first equation has its counterpart in 
\begin{equation}
\sum_{n=1}^{\infty} p_{n}(x,t) = cL f(x,t)\,\,, 
\end{equation}  
where $cL$ is the mean number of particles initially distributed in $(0,L)$. The normalization condition undergoes more significant changes. When $N$ is random, it is no longer sure that the $n$-th particle is initially located in $(0,L)$. Therefore we have 
\begin{equation} 
\label{norm}
\int_{0}^{L}\!\!\! {\rm d} x\, p_{n}(x,0)=1-p(N<n)\,\,, 
\end{equation}  
where $p(N<n)$ stands for the probability that the initial number of particles is less than $n$.  

The probability that, at the time $t$, the $n$-th tagged particle is in $(0,L)$, 
\begin{equation}
S_{n}(t) = \int_{0}^{L}\!\!\! {\rm d} x\, p_{n}(x,t)\,\,, 
\end{equation}
reads (cf. Appendix \ref{AppendixS})
\begin{eqnarray} 
\label{Sn}
S_{n}(t) &=&  \frac{1}{(n-1)!} \sum_{k=0}^{\infty}\frac{(-1)^{k}}{n+k} \frac{(cL)^{n+k}}{k!} \times \\ 
\nonumber
& & \times \left[ \left(F(L,t)\right)^{n+k}-  \left(F(0,t)\right)^{n+k} \right]\,\,. 
\end{eqnarray}
Notice that the survival probability (\ref{Sn}) can also be obtained from Eq.\ (\ref{SnN}) as follows: 
\begin{equation}
\label{SnSUM}
S_{n}(t) = \sum_{N=n}^{\infty}  S_{n:N}(t)P(N)\,\,.
\end{equation}

The results derived in the present section hold true for any of the two types of boundary conditions. Let us now treat individual settings in a more detail. 

\section{The left boundary is reflecting the right boundary is absorbing}
\label{Transmission}
\subsection{Single noninteracting particle}

The pdf of the single noninteracting particle satisfies the diffusion equation (\ref{Diff})
subject to the absorbing boundary condition at the origin, $f_{\rm T}(0,t)=0$, the reflecting boundary condition at $x=L$, $\partial f_{\rm T}/\partial x |_{x=L}= 0$, and the initial condition $f_{\rm T}(x,0)=1/L$, $x\in (0,L)$. The solution of this initial-boundary value problem can be expressed as the eigenfunction expansion  
\begin{eqnarray} 
\nonumber 
f_{\rm T}(x,t) &=& \frac{4}{\pi L} \sum_{k=0}^{\infty} \frac{(-1)^{k}}{2k+1} \cos\!\left[\left(1-\frac{x}{L}\right)\frac{\pi}{2}  (2k\!+\!1)\right] \times \\
& & \times \exp\!\left\{-\left[\frac{\pi}{2L}(2k+1)\right]^{2} D t\right\}
\,\,. 
\end{eqnarray} 
Each term of the above series decays exponentially in time with a decay rate being higher for higher values of $k$.
The survival probability of the single noninteracting particle,  
\begin{equation}
\label{STdef}
S_{\rm T}(t) = \int_{0}^{L}\!\!\! {\rm d} x\, f_{\rm T}(x,t)\,\,,
\end{equation}
has an asymptotic representation 
\begin{equation}
\label{ST}
S_{\rm T}(t) \sim \frac{8}{\pi^{2}} \, 
 \exp\!\!\left[-\left(\frac{\pi}{2L}\right)^{2}\!\! D t\right]\,\,.
\end{equation}

\subsection{Fixed $N$}

Let us choose an arbitrary fixed $x$, $x \in (0,L)$. At a given time $t$, the probability that the noninteracting particle is found to the left of $x$, $F(x,t)$, is given by the sum of probabilities of two disjoint events. First, the probability that the particle has already been trapped before $t$. Second, the probability that, at the time $t$, the particle is found within the interval $(0,x)$. Since the right boundary is reflecting, we get
\begin{equation}  
\label{CDF1}
F(x,t) = \left( 1-S_{\rm T}(t) \right) + \int_{0}^{x}\!\!\! {\rm d} x' f_{\rm T}(x',t)\,\,.
\end{equation}
It follows that  
\begin{equation}
\label{FT}
1-F(0,t) = S_{\rm T}(t)\,\,,\qquad 1-F(L,t) = 0 \,\,.
\end{equation}
We introduce the two results (\ref{FT}) into Eq.\ (\ref{SnN}) thus obtaining the exact expression for the tagged particle survival probability. We have 
\begin{eqnarray}
S_{n:N}^{\rm T}(t) &=&  n \binom{N}{n}  \times \\ 
\nonumber
& &\!\!\!\!\!\!\!\!\!\!\!\!\!\!
 \times \sum_{k=0}^{n-1}\frac{(-1)^{k}}{N-n+k+1} 
\binom{n-1}{k} 
 \left(S_{\rm T}(t)\right)^{N-n+k+1}. 
\end{eqnarray}
Each summand of the above sum decays exponentially with time (cf.\ Eq.\ (\ref{ST})). The leading term in a long-time limit is that with $k=0$. It reads 
\begin{equation}
\label{SnNT}
S_{n:N}^{\rm T}(t) \sim 
 \frac{n}{N-n+1} \binom{N}{n} 
 \left(S_{\rm T}(t)\right)^{N-n+1}\,\,. 
\end{equation}
Substituting (\ref{ST}) into the above result we obtain the explicit form of the tagged particle asymptotic survival probability: 
\begin{eqnarray}
\label{SnNTe}
S_{n:N}^{\rm T}(t)  \sim C_{n:N}^{\rm T} \exp\!\!\left[-(N-n+1)\left(\frac{\pi}{2L}\right)^{2}\!\! D t\right] 
 \,\,,
\end{eqnarray}
where the prefactor reads
\begin{equation}
C_{n:N}^{\rm T} =  \frac{n}{N-n+1} \binom{N}{n} 
 \left( \frac{8}{\pi^{2}}\right)^{N-n+1}\,\,. 
\end{equation}

Notice that the asymptotic tagged particle survival probability decays exponentially with time, the decay rate being determined by the number of particles located between the tagged particle and the hard wall, ($N-n$). In particular, the survival probability of the right-most particle ($n=N$) decays exactly with the same rate as that of the single-diffusing particle (cf.\ Eq.\ (\ref{ST})). In this regard the result (\ref{SnNT}) is similar to the one obtained in Ref.\ \onlinecite{RC2012} for a semi-infinite interval with the absorbing boundary at the origin (where the single-particle survival probability decays as a power law $t^{-1/2}$).

\subsection{Fixed initial concentration of particles}

For a given $N$, Eqs.\ (\ref{SnNT}), (\ref{SnNTe}) demonstrate an effect of entropic repulsion among the particles: the more particles are located between the tagged particle and the reflecting boundary the higher the decay rate of the tagged particle survival probability. Hence, in the present setting (when $c$ is given but $N$ is random), one could expect that the decay rate of the tagged particle survival probability is controlled by an average number of particles located between the tagged particle and the reflecting boundary. However, this is not the case. 

According to Eq. (\ref{SnSUM}), the survival probability of the $n$-th particle is given by
\begin{equation} 
\label{SnTSUM}
S_{n}^{\rm T}(t) = \sum_{N=n}^{\infty} S_{n:N}^{\rm T}(t) P(N) \,\,.
\end{equation} 
In the long-time limit, only the first term ($N=n$) of (\ref{SnTSUM}) contributes significantly to the sum (cf.\ Eqs.\ (\ref{SnNT}), (\ref{SnNTe})). Thus the asymptotic survival probability reads 
\begin{equation}
\label{SnT} 
S_{n}^{\rm T}(t) \sim cL \frac{(cL)^{n-1}}{(n-1)!}\, {\rm e}^{-cL} S_{\rm T}(t)\,\,. 
\end{equation} 
Hence the long-time behavior of the survival probability $S_{n}^{\rm T}(t)$ is dominated by those cases when the tagged particle is initially the right-most one ($N=n$). 

\section{Both boundaries are absorbing}
\label{Absorption}
\subsection{Single noninteracting particle}

Let us now turn to the second generic case when the both boundaries are absorbing. Now, the diffusion equation (\ref{Diff})
is supplemented by the boundary conditions $f_{\rm A}(0,t)=f_{\rm A}(L,t)=0$, and by the initial condition $f_{\rm A}(x,0)=1/L$, $x\in (0,L)$. The solution reads  
\begin{equation}
f_{\rm A}(x,t) = \frac{4}{\pi L} \sum_{k=0}^{\infty}\frac{\sin\!\left(\pi x (2k\!+\!1)/L\right)}{2k+1} \, 
{\rm e}^{-\left[\pi (2k+1\right)/L]^{2} D t}
\,\,.
\end{equation}

The survival probability of the single noninteracting particle, $S_{\rm A}(t)$, is given by a spatial integral of pdf $f_{\rm A}(x,t)$ over the interval (cf. Eq.\ (\ref{STdef})). The asymptotics of $S_{\rm A}(t)$ reads
\begin{equation}
\label{SA}
S_{\rm A}(t) \sim \frac{8}{\pi^{2}} \, 
 \exp\!\!\left[-\left(\frac{\pi}{L}\right)^{2}\!\! D t\right]\,\,.
\end{equation}

\subsection{Fixed $N$}
\label{AbsorptionN}

Since $f_{\rm A}(x,0)$ is uniform, the probability that the noninteracting particle is trapped by the left boundary before $t$ equals $ \left( 1-S_{\rm A}(t) \right)/2$. Hence the probability $F(x,t)$ to find the noninteracting particle to the left of $x$, $x\in (0,L)$, is given by (compare this result with Eq.\ (\ref{CDF1}))
\begin{equation}
\label{CDF2}
F(x,t) = \frac{1}{2} \left( 1-S_{\rm A}(t) \right) + \int_{0}^{x}\!\!\! {\rm d} x' f_{\rm A}(x',t)\,\,.
\end{equation}
The above formula yields
\begin{equation}
\label{FA0}
1-F(0,t) =  \frac{1+S_{\rm A}(t)}{2}\,\,,\quad 1-F(L,t) =  \frac{1-S_{\rm A}(t)}{2} \,\,.
\end{equation}

In order to obtain a tagged particle survival probability $S_{n:N}^{\rm A}(t)$ we substitute expressions (\ref{FA0}) into Eq.\ (\ref{SnN}). In the long-time limit, the leading term of the resulting sum reads  
\begin{equation}
\label{SnNA}
S_{n:N}^{\rm A}(t) \sim 
 \frac{n}{2^{N-1}}  \binom{N}{n} 
 S_{\rm A}(t)\,\,. 
\end{equation} 
The asymptotic survival probability (\ref{SnNA}) decays exponentially with time, the decay rate being independent of $n$. The hard-core interaction manifests itself only through the $n$-dependent prefactor. Moreover, the asymptotic decay rate of $S_{n:N}^{\rm A}(t)$ is the same as that for the single noninteracting particle.  This can be understood on physical grounds. Due to the mutual entropic repulsion among the particles, it is highly probable that the long-lived tagged particle eventually remains alone in the interval. Therefore its long-time dynamics should resemble that of the noninteracting particle. 

\subsection{Fixed initial concentration of particles}

Consider now fixed $c$ and random $N$ case. The survival probability of the $n$-th particle, $S_{n}^{\rm A}(t)$,  is obtained from $S_{n:N}^{\rm A}(t)$ by performing the summation (cf. Eq.\ (\ref{SnSUM}))
\begin{equation}
\label{SnASUM}
S_{n}^{\rm A}(t) = \sum_{N=n}^{\infty} S_{n:N}^{\rm A}(t) P(N) \,\,,
\end{equation}
where $P(N)$ is the Poisson distribution (\ref{Poisson}). In the long-time limit, \emph{all terms of the above sum are proportional to} $S_{\rm A}(t)$ (cf. Eq.\ (\ref{SnNA})). The asymptotic survival probability is given by
\begin{equation} 
\label{SnA}
S_{n}^{\rm A}(t) \sim cL \frac{(cL/2)^{n-1}}{(n-1)!}\, {\rm e}^{-cL/2} S_{\rm A}(t)\,\,.
\end{equation}
The above result is formally similar with Eq.\ (\ref{SnT}). However, the physical background of both asymptotics is rather different. 
In contrast to Eq.\ (\ref{SnT}),  Eq.\ (\ref{SnA}) represents the overall contribution of all possible initial particle numbers for which $n\leq N$ (while Eq.\ (\ref{SnT}) gives only the first term of the sum (\ref{SnTSUM})).

\subsection{Random interval length}
\label{randomintlength}

Finally, let us derive the asymptotic tagged particle survival probability when $L$ is drawn from the exponential distribution with the mean $1/\lambda$. This corresponds to the model of diffusion  on the real line with randomly distributed perfectly absorbing traps. The concentration of traps is uniform and equals $\lambda$. We assume that the initial concentration of particles  $c$ is also uniform. Actually, for non-interacting particles, there exists a large body of literature on the subject, see e.g. Refs.\ \onlinecite{Havlin, HavlinAdvances, Anlauf, Grassberger, WeissHavlin, Yuste, RednerKang} and references therein. 

The average of the survival probability $S_{n}^{\rm A}(t)$ over the probability distribution of the interval lengths is given by 
\begin{equation}
\label{SnAaveragedDef}
\bar{S}_{n}^{\rm A}(t) = \int_{0}^{\infty} \!\!\!{\rm d}L\,  \lambda {\rm e}^{- \lambda L} S_{n}^{\rm A}(t) \,\,.
\end{equation} 
We substitute the asymptotics (\ref{SnA})  into the above integral. After that the integral is evaluated by a saddle-point method (see Appendix \ref{AppendixL}). The resulting long-time asymptotics of $\bar{S}_{n}^{\rm A}(t)$ reads
\begin{equation} 
\label{SnAaveragedAsy}
\bar{S}_{n}^{\rm A}(t) \sim C (Dt)^{(2n+1)/6} \!
\exp\!\left\{- \frac{3}{2} \left[2 \pi^{2}\!\! \left(\lambda\!+\!\frac{c}{2}\right)^{2}\!\! Dt \right]^{1/3}\! \right\}\,,
\end{equation}
where the prefactor is 
\begin{equation} 
C= \sqrt{\frac{2\pi}{3}}
\frac{16\lambda}{\pi^{2}}
\frac{(c/2)^{n}}{(n-1)!}
\frac{(2 \pi^{2})^{(2n+1)/6}}{(\lambda+c/2)^{(n+2)/3}} \,\,.
\end{equation}

As it is for the single noninteracting particle,\cite{Havlin} the decay of the survival probability is slower than its decay when the interval length is fixed. The stretched exponential relaxation in Eq.\ (\ref{SnAaveragedAsy}) results from very large (and extremely rare) interval length $L$ which enhance the asymptotic long-time survival probability.\cite{Grassberger}  A remarkable feature, which is not presented in a noninteracting model, is the dependence of the factor in the exponential on $c$. That is on the concentration of particles and not only on the concentration of traps $\lambda$. This dependence is the consequence of the entropic repulsion among particles -- the higher the initial concentration the stronger the effective force that pushes the tagged particle into the trap. 

\section{Concluding remarks}
\label{concludingremarks}

The main results of the paper are the asymptotics (\ref{SnNTe}), (\ref{SnT}), (\ref{SnNA}), (\ref{SnA}), (\ref{SnAaveragedAsy}). 
When the right boundary is reflecting, the particles located between the reflecting boundary and the $n$-th tagged particle create an effective (entropic) repulsive force which shortens the time spent by the tagged particle in the interval. This is reflected in the dependence of the decay rate of asymptotic tagged-particle survival probability (\ref{SnNTe}) on $N-n$. If the initial concentration is given instead of the precise initial number of particles, then Eq.\ (\ref{SnT}) yields the asymptotic survival probability averaged over all possible initial particle numbers. In fact, Eq.\ (\ref{SnT}) represents only the first term of the sum (\ref{SnTSUM}). Thus the long-time properties are controlled by the initial condition in which the $n$-th particle is the right-most one.

The situation is rather different when both boundaries are absorbing. For a given $N$, the long-time regime is governed by the situations when the tagged particle remains alone in the interval. Then its dynamics resembles that of the noninteracting particle. The only evidence of the hard-core interaction is the pre-exponential factor of the asymptotic survival probability (\ref{SnNA}). For a given $c$, all terms of the sum (\ref{SnASUM}) are asymptotically proportional to the survival probability of the single noninteracting particle $S_{\rm A}(t)$. Hence all initial conditions with all possible initial particle numbers contribute to the asymptotics (\ref{SnNA}). When $L$ is random (and $c$ is given), the tagged-particle asymptotic survival probability (\ref{SnAaveragedAsy}) exhibits a stretched-exponential decay. This is in parallel to what is known for the case without interaction.\cite{Havlin} A remarkable consequence of the hard-core interaction is the dependence of the asymptotic survival probability (\ref{SnAaveragedAsy}) on the combination $\lambda + c/2$. That is, not only the concentration of traps $\lambda$ shortens the  lifetime of the tagged particle but also the initial particle concentration $c$. 

Finally, notice that the results obtained in the present paper (and in Ref.\ \onlinecite{RC2012}) are different as compared to the result obtained in Ref.\ \onlinecite{FPTSFD}. The difference stems from the fact that in the present study (and in Ref.\ \onlinecite{RC2012}) all particles are absorbed at the boundary whereas in Ref.\ \onlinecite{FPTSFD} only the tracer particle ``senses'' the presence of the absorbing boundary.

\begin{acknowledgments}
This work was supported by the grant SVV-2013-267305, and by the Charles University Grant Agency (projects No.\ 301311, and No.\ 143610).
\end{acknowledgments}

\appendix
\section{Derivation of $S_{n:N}(t)$ and $S_{n}(t)$}
\label{AppendixS} 

Notice that $f(x,t)= \partial F(x,t)/\partial x$. By definition we have (for the sake of brevity the arguments of $F(x,t)$ are omitted)
\begin{equation} 
S_{n:N}(t) = n \binom{N}{n} \int_{0}^{L}\!\!\! {\rm d} x \, 
 F^{n-1} \left(1-F\right)^{N-n}\frac{\partial F}{\partial x}\,\, .
\end{equation}
Next we expand the term $\left[1-(1-F)\right]^{n-1}$ according to the binomial theorem. We get
\begin{eqnarray}
S_{n:N}(t) &=&  n \binom{N}{n} \sum_{k=0}^{n-1} (-1)^{k} \binom{n-1}{k} \times \\
\nonumber
& &\times \frac{(-1)}{N-n+k+1}
\int_{0}^{L}\!\!\! {\rm d} x \frac{\partial}{\partial x} \, 
 \left(1-F\right)^{N-n+k+1}\,\, ,
\end{eqnarray} 
which, after performing the required integration, yields Eq.\ (\ref{SnN}).

Derivation of the survival probability (\ref{Sn}) requires similar steps. First, we expand the exponential in Eq.\ (\ref{pn}) into the power series, then, using the fact that $f(x,t)= \partial F(x,t)/\partial x$, the integration over $x$ is carried out which results in Eq.\ (\ref{Sn}). 

\section{Laplace's approximation} 
\label{AppendixL}

To keep the paper self-contained let us present an asymptotics of the integral arising in Sec.\ \ref{randomintlength} \cite{Agmon}
\begin{equation} 
I(t) = \int_{0}^{\infty}\!\!\!\!{\rm d}L\,\, L^{n} \exp\!\left[- \alpha L - \frac{\beta t}{L^2} \right]\,\,,\quad \alpha, \beta >0\,\,.
\end{equation}
The substitution $L= t^{1/3} x$ brings us to the expression
\begin{equation}
I(t) = t^{(n+1)/3} \int_{0}^{\infty}\!\!\!\!{\rm d}x\,\, x^{n} \exp\left[- t^{1/3}\left( \alpha x + \frac{\beta}{x^2} \right) \right]\,\,.
\end{equation}
The above integral is approximated by the Laplace's method.\cite{Henrici2} 
 The approximation of the integral reads
\begin{equation}
I(t) \sim C_{n} \, t^{(2n+1)/6}  \exp\!\left[-\frac{3}{2}\left(2\beta \alpha^{2} t\right)^{1/3} \right]\,\,,
\end{equation}
where
\begin{equation}
C_{n} = \sqrt{\frac{2\pi}{3}} \frac{(2\beta)^{(2n+1)/6}}{\alpha^{(n+2)/3}}\,\,.
\end{equation}


\bibliography{references}

\end{document}